 \documentclass[final,3p,times,twocolumn]{elsarticle}

 \usepackage{graphicx}

\usepackage{amssymb}

\journal{Physics Letters B}

\begin{document}

\begin{frontmatter}



\title{Evidence of $\alpha$ particle condensation in $^{12}$C and $^{16}$O 
 and Nambu-Goldstone boson}


\author[rvt]{S. Ohkubo} 
\ead{ohkubo@yukawa.kyoto-u.ac.jp}


\address[rvt]{University of Kochi , Kochi 780-8515, Japan and\\
Research Center for Nuclear Physics, Osaka University, Ibaraki, 
Osaka 567-0047, Japan}

\begin{abstract}
\par
It is suggested that direct evidence of Bose-Einstein condensation of $\alpha$ particles
is obtained by observing a phase mode (Nambu-Goldstone boson) with long
 wavelength even when characteristic features such as  superfluidity  is diffucult  
 to observe. For the 7.65 MeV  $0_2^+$ Hoyle state in  $^{12}$C and  15.1 MeV $0^+$ state in 
 $^{16}$O, which are candidates for an $\alpha$ particle condensate, it is suggested that  
 the emergent band head  $0^+$  state  of the $K=0_2^+$  rotational  band with a very large moment 
of inertia is considered to be a   Nambu-Goldstone boson.
\end{abstract}

\begin{keyword}
$\alpha$ particle condensation  \sep Nambu-Goldstone boson 
\sep spontaneous symmetry breaking \sep $\alpha$ cluster structure 
 \sep $^{12}$C; $^{16}$O  
\PACS  21.60.Gx   \sep 27.20.+n  \sep 03.75.Nt


\end{keyword}
\end{frontmatter}


 The  $0_2^+$ state in $^{12}$C at  excitation energy $E_x$=7.65 MeV, the  Hoyle state, is a key 
 state for  nucleosynthesis, the evolution of stars    and the emergence of life. 
The existence of this resonant  state  just above the  $\alpha$ threshold was  predicted 
by Hoyle to explain the enhanced triple $\alpha$ reactions needed to understand 
the nucleosynthesis of $^{12}$C. In the last 60 years, the understanding of the Hoyle state 
has been deepened by  the study of the $\alpha$ cluster structure
of $^{12}$C.  However,  its interpretation has changed considerably.
Morinaga \cite{Morinaga1956} proposed that the Hoyle state is a band head state with 
a linear chain  structure of three $\alpha$
 particles. The three $\alpha$ structure of $^{12}$C was most thoroughly investigated by 
 Uegaki  et al. \cite{Uegaki1977} in their  pioneering work, 
 which showed that the Hoyle state has a dilute structure in a new  
``$\alpha$-boson gas phase''   and clarified  the systematic  existence of a
 ``new phase'' of the three $\alpha$ particles above the $\alpha$ threshold.

 Inspired by  the observation of Bose-Einstein condensation of cold  atomic gas 
clusters, much attention has been paid to investigate whether Bose-Einstein 
 condensation of 
$\alpha$ particles occurs in light nuclei where $\alpha$ cluster structure  widely exists.
The three $\alpha$ particle system of $^{12}$C, especially the dilute properties of the Hoyle
state due to $\alpha$ particle condensation,    has been studied
 extensively by many authors 
\cite{Tohsaki2001,Funaki2003,Ohkubo2004,Yamada2005,Chernykh2007,Ohkubo2007,Danilov2009,
Raduta2011,Manfredi2012}. 
 It is now evident theoretically and experimentally that the Hoyle state has a larger radius
  compared with the ground state and has a dilute matter distribution.

 In order to get  direct experimental  evidence that the  Hoyle state is 
a  condensate,  Raduta   et al.  \cite{Raduta2011} observed   three
 $\alpha$ particles  with the same energy
 emitted simultaneously from the Hoyle state and  suggested that
 the three $\alpha$ particles were sitting in the lowest $0s$ state.
On the other hand, in a similar coincidence experiment  Manfredi  et al.
\cite{Manfredi2012} skeptically set an
  upper limit of  a probability of 0.45\%,  which
 confirms the previous result by   Freer  et al. \cite{Freer1994}.
 In spite of much effort, firm experimental evidence of Bose-Einstein condensation such 
as superfluidity  or a vortex  has not been found  for the Hoyle state.

The purpose of this paper is to suggest that evidence of $\alpha$ particle 
condensation can be provided by the observation of the  Nambu-Goldstone (NG)
 boson caused by the spontaneous symmetry breaking of the global phase symmetry.
 We suggest that for $^{16}$O  a bandhead $0^+$ state predicted at 16.6 MeV that accompanies 
a $K=0_2^+$ band with  very large moment inertia  is a NG boson collective zero mode
 associated with the condensation of   four $\alpha$  particles of the  0$^+$ state at 
15.1 MeV.
 We also suggest in $^{12}$C that a $0_3^+$ state and a rotational band built on it should emerge 
slightly above  
 the Hoyle state as a NG boson if the Hoyle state is a condensate of three 
 $\alpha$ particles. We show an experimental candidate   for the NG boson state 
in $^{16}$O and $^{12}$C.

If the concept  of $\alpha$ particle condensation persists in  nuclei, 
the four $\alpha$ particle system of $^{16}$O is  another good candidate.
Tohsaki et al. \cite{Tohsaki2001} conjectured   that the $0_3^+$ 
 state at 11.26 MeV in $^{16}$O, which is located 3.18 MeV below the four $\alpha$
 threshold energy, may be  an $\alpha$ particle condensate  because it is well 
represented by a Bose wave function with a dilute  four $\alpha$ particle structure.
Funaki et al. \cite{Funaki2008C} investigated  0$^+$ states in $^{16}$O using a 
four $\alpha$ cluster model in the bound state approximation and   suggested that
 the 0$^+$ state at 15.1 MeV is probably  an $\alpha$ condensed state.
  Ohkubo and  Hirabayashi \cite{Ohkubo2010} investigated not only the 0$^+$ states but also the 
so-called four $\alpha$ linear chain states with higher
 spins \cite{Chevallier1967,Freer1995}
from the viewpoint of the unified description of quasi-bound states and
 $\alpha$ scattering by analyzing the $\alpha$+$^{12}$C elastic scattering and 
$\alpha$+$^{12}$C(0$^+_2$) inelastic scattering using  a coupled channel method.
They showed that the so-called four $\alpha$  linear chain states can be understood 
as a  $K=0_2^+$ band  (Fig.~1) with the $\alpha$+$^{12}$C(0$^+_2$) configuration.  
They  suggested that the observed 15.1 MeV $0^+$ state is a candidate for  the 
 four $\alpha$ particle  condensate  in a  completely different quantum state under 
Bose statistics.

\begin{figure}[tbh]
   \includegraphics[width=8.cm]{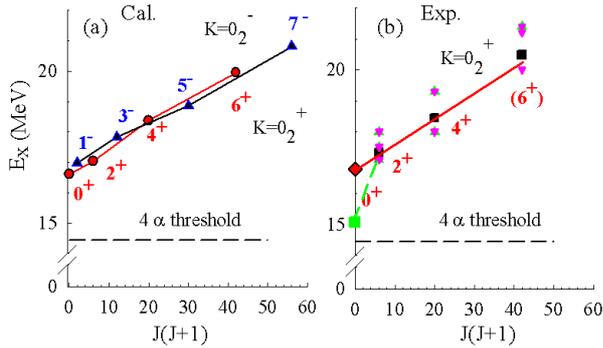}
 \protect\caption{\label{fig.1} {
(a) The calculated     $K=0^+_2$ and $K=0^-_2$
  rotational bands with the $\alpha$+$^{12}$C($0^+_2$)  cluster  structure  in  
 $^{16}$O  near the four $\alpha$ threshold \cite{Ohkubo2010}. 
   (b) The experimental   rotational band  states with $\alpha$ cluster structure
 taken from  Ref.\cite{Ohkubo2010,Itoh2012}.
The centroid of each of the spin states is  shown by a black square.  
 The candidate 15.1 MeV $0^+$ state  of four $\alpha$ condensate is displayed by 
a green square.  The lines are   to guide the eye. 
}
}
 \end{figure}%

 If the 15.1 MeV 0$^+$ state is a special state of  a Bose-Einstein condensate 
 of  four $\alpha$ particles, 
  the state can be  described by a macroscopic wave function 
 in the Ginzburg-Landau theory,  
$\Psi=$$\mid$$\Psi$$\mid$e$^{i\theta}$ where $\mid$$\Psi$$\mid$ is the 
order parameter.
In a condensate,  continuous global phase symmetry in gauge space is spontaneously 
broken  with $\mid$$\Psi$$\mid$$\ne0$ in the NG phase and
 two  collective 
modes appear, i.e., a massless (zero energy with a infinite wavelength)
 phase ($\theta$) mode  (NG boson)  and a finite mass amplitude ($\mid$$\Psi$$\mid$) mode 
  (Higgs boson)     \cite{Nambu1961}. 
Spontaneous symmetry breaking is ubiquitous in physical systems and in nature \cite{Watanabe2012}.
For example, for  infinite systems, in  superconductors the NG mode (in which the NG boson has been 
eaten by the plasmon  \cite{Kadowaki1998}) and the  Higgs mode \cite{Littlewood1982} have been observed.  
As seen in Fig.~2, in particle  physics the  massless pion and the amplitude mode 
$\sigma$ meson emerge as  a collective mode from the vacuum 
due to the chiral condensation of the vacuum \cite{Nambu1961}. 
We note that a band (Regge
 trajectory) of the states, in which orbital angular momentum 
 of the relative motion  between a quark and an anti-quark is excited, 
is built on top of the NG boson pion.  The spectra including the radially excited states are
 explained  well by  the quark model \cite{Ebert2009}. 
For  finite systems a Nambu-Goldstone boson as well as a Higgs mode boson  have been observed 
in superfluid nuclei as  a pairing rotation and    a pairing vibration, respectively
 \cite{Broglia1973}.
For Bose-Einstein condensed cold atom gases with finite number of particles in a finite volume, 
the existence of a Nambu-Goldstone boson has
 been shown theoretically from the Ward-Takahasi identity in quantum  field theory
 \cite{Enomoto2006,Mine2006}.

The appearance of the $K=0_2^+$ band in Fig.~1(a) predicted in Ref.\cite{Ohkubo2010}
 just above the 15.1 MeV 0$^+$  state    may be 
 understood from the viewpoint of spontaneous symmetry breaking 
 as follows: If the 15.1 MeV 0$^+$ state is a condensate,
a NG boson zero energy collective mode should appear. 
Considering that (1) the 0$^+$ at 16.6 MeV is a first $0^+$ state above the condensed 
vacuum (15.1 MeV  $0^+$ state) with  the very small excitation energy of  1.5 MeV
(almost massless) and (2) the very large moment of inertia of the band built on it,
which is due to a collective motion with the largest dimension that  the system allows 
(long wavelength),  this state may be regarded as the emergence of 
 the NG boson caused by the spontaneous symmetry breaking 
of the global phase due to the four $\alpha$ particle condensation.
This NG state   is a collective state with enlarged structure  of four $\alpha$
 particles with dilute property   due to its 
long wavelength nature, i.e. deformed orthogonal to 
 the spherical vacuum 15.1 MeV 0$^+$  in the $0s$ states.

In fact, in Ref.\cite{Ohkubo2010} it was shown that the  
  0$^+$ resonant state predicted at 16.6 MeV has
a loosely coupled $\alpha$+$^{12}$C($0_2^+$) cluster structure with large deformation. 
As shown in Fig.~1 (a), their calculated resonances with spin 2$^+$, $4^+$ and $6^+$  are located 
on the J(J+1) line forming a rotational band.
The predicted negative parity band $K=0_2^-$ \cite{Ohkubo2010} is considered as 
a parity-doublet partner of the   $K=0_2^+$ band  with  almost the same moment  of inertia. 
The very large moment inertia of the $K=0_2^+$ band stems from the NG boson nature
 of the band head  0$^+$ state with a massless long wave  collective motion. 
It is noted that the calculated resonances of the $K=0_2^+$  band correspond well
 with the experimental $\alpha$ cluster states observed in the four $\alpha$ coincidence 
experiments \cite{Chevallier1967,Freer1995}.

Before proceeding to the experimental candidate for  the NG boson $0^+$ state, we
would like to emphasize that the 15.1 MeV  0$^+$ state should  not considered
 as a member  state of the $K=0_2^+$ band built on the NG 0$^+$ state. 
 The intrinsic structure of the 15.1 MeV 0$^+$ state is 
quite different in nature  from the  deformed NG boson  $0^+$ state and
 its associated rotational band states.
 The NG boson state is a logical consequence of the existence of the  15.1 MeV
   state that spontaneously violates the global phase symmetry due to condensation.  
While the 15.1 MeV 0$^+$ state is  spherical  
 with Bose statistics in the $0s$ state,  the $K=0_2^+$ rotational 
band states are deformed in  a local condensate state in which the three 
$\alpha$ particles are condensed forming the Hoyle state \cite{Ohkubo2010}. 
If we connect the  15.1 MeV 0$^+$ state and the 2$^+$ state (Fig.~1(b)),
the estimated moment of inertia  appears  drastically 
reduced  from  that of the  $K=0_2^+$ rotational band as discussed in Ref.\cite{Ohkubo2010}.

\begin{figure}[t]
   \includegraphics[width=6.0cm]{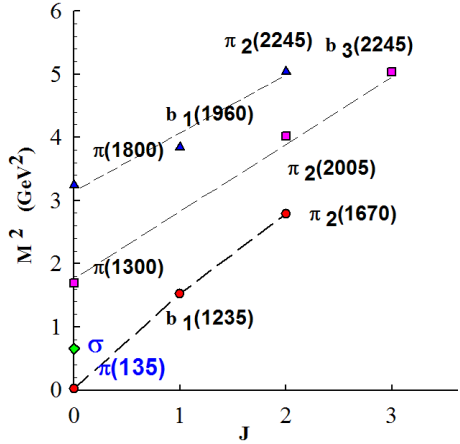}
 \protect\caption{\label{fig.2} {The experimental spectra 
of the NG boson pion, and the  orbital and radial 
excited states of the pion are displayed together with the Higgs boson $\sigma$ 
meson.  The lines  are to guide the eye. Data are from Ref.\cite{Amsler2008}. 
 }
}
 \end{figure}

 According to our interpretation, the existence of the NG boson 0$^+$ 
 state predicted at 16.6 MeV is logically  very important. However, no 0$^+$ state
 has been  reported around here in the  literature. 
It may not  be easy to  observe the  NG boson because of a large decay 
width of  more than 1 MeV.
Very recently Itoh  et al. reported  \cite{Itoh2012}
that they  have  observed   new broad resonant  0$^+$ states at 16.8 MeV
and 18.8 MeV. The  16.8 MeV state with an $\alpha$ cluster structure of  a width of  about  1 MeV 
(a diamond in Fig.~1(b)) 
agrees precisely  with  the    theoretically predicted energy 16.6 MeV  and the
$\alpha$ width  1.1 MeV (Fig.~1(a)) \cite{Ohkubo2010}.
This   strongly  supports  the present interpretation.
The 18.8 MeV $0^+$ state may correspond to the $0_4^+$ state around 10 MeV in Fig.~3.

Now we come to the  $\alpha$ particle condensation of three $\alpha$ particles in $^{12}$C.
According to the above interpretation, the understanding of the $K=0_2^+$  rotational
 band built on the Hoyle state  
is drastically changed. Morinaga \cite{Morinaga1956} interpreted that the Hoyle state and
 the 10.3 MeV 2$^+$ state
form a rotational band with a very large moment of inertia, which leads to the three $\alpha$
 linear chain model.   Recently $2_2^+$ (9.6 MeV) \cite{Zimmerman2011,Itoh2011,Freer2012}
and $4_1^+$ (13.3 MeV) \cite{Freer2011} states above the Hoyle state
 have been observed.   The $2_2^+$ state  has 
been considered to be a rotational band state built on the Hoyle state 
\cite{Kurokawa2005}.
If the Hoyle state is a condensate of three $\alpha$ particles,
  a NG boson zero energy collective mode should appear 
just above it.
The NG boson is massless in principle, however, in nature its energy  is not always  exactly 
zero   as seen in the pion case. If the NG state has  zero energy, the   $2_2^+$ (9.63 MeV) 
and $4_1^+$ (13.3 MeV) above  stand on the Hoyle state. In this case it is unlikely 
 that the $2_2^+$ and $4_1^+$ states are rotational member states because the condensate
 Hoyle state is spherical with the three $\alpha$ particles sitting in the 0s state.
On the other hand, if the NG boson mode energy is not exactly zero as in the
 case of $^{16}$O, another $0^+$ state emerges just above the Hoyle state. 
Because the NG $0^+$ state is not spherical, a rotational band should appear.
The observed $2_2^+$ (9.63 MeV) and $4_1^+$ (13.3 MeV) states are considered to be 
a member of this rotational band.  The band head NG boson $0^+$ state and the   $2^+$ and $4^+$ 
states  should have the same $\alpha$ cluster configuration in nature, mostly a loosely coupled 
$\alpha$+$^8$Be configuration  created by lifting an $\alpha$ particle from the
 vacuum (Hoyle state). As shown in Fig.~3, this is an analog  
 of the loosely coupled $\alpha$+$^{12}$C($0_2^+$) structure built on the 
NG boson $0^+$ state predicted at 16.6 MeV in $^{16}$O. The intraband $B(E2)$ 
transitions should  be very large  if measured because of a large deformation, i.e. large moment
 of inertia. On the other hand the $B(E2)$ transition from the $2_2^+$ to the Hoyle
 state should  be much smaller than that from the intraband
 $B(E2$:$2_2^+$$\rightarrow$0$_3^+$).

The results of the {\it ab initio} calculation of $^{12}$C in the bound state approximation 
 in Ref.\cite{Kanada2007} seems to support the present interpretation theoretically.
The calculated  $0_3^+$, $2_2^+$  and $4_1^+$ states have the  $\alpha$+$^8$Be
 intrinsic configuration and 
 the intraband transition  $B(E2)$ values are very large, 600 and 310 (e$^2$fm$^4$) for
the transitions  $4_1^+$$\rightarrow$$2_2^+$,   2$_2^+$$\rightarrow$0$_3^+$, respectively.
 On the other hand the $B(E2)$ transition from the  $0_3^+$ to the Hoyle state is one third, 
 100 (e$^2$fm$^4$). Kurokawa and Kat\=o emphasized  \cite{Kurokawa2005,Kurokawa2007} 
the importance of treating the  unbound resonance states in this energy region  carefully 
under the  correct three-body resonance condition
and found   broad resonance states, $0_3^+$  at  $E_x$=8.95 MeV with $\Gamma$=1.48 MeV
 (only 1.66 MeV  above the $\alpha$ threshold) and  $0_4^+$  at  $E_x$=11.87 MeV with 
$\Gamma$=1.1 MeV Ref.\cite{Kanada2007}. They suggested that this $0_3^+$ state may be 
 a higher nodal state, in which
radial motion between $^8$Be and the $\alpha$ particle is  excited \cite{Kurokawa2007}.
We note this $\alpha$ cluster $0_3^+$ state is located below the 
newly observed $2_2^+$ (9.63 MeV) state and may correspond to our NG boson
 $0^+$ state  because (1) the excitation energy 1.28 MeV from the  Hoyle 
state is small,  which is similar  to the 1.5 MeV of the $^{16}$O case (massless) and (2) this $0_3^+$ state
forms a $K=0_2^+$ rotational band with the $2_2^+$ (9.63 MeV) and the $4_1^+$ (13.3 MeV)
 states with a large moment of inertia similar to the $^{16}$O case (long wavelength).

\begin{figure}[t]
   \includegraphics[width=7.0cm]{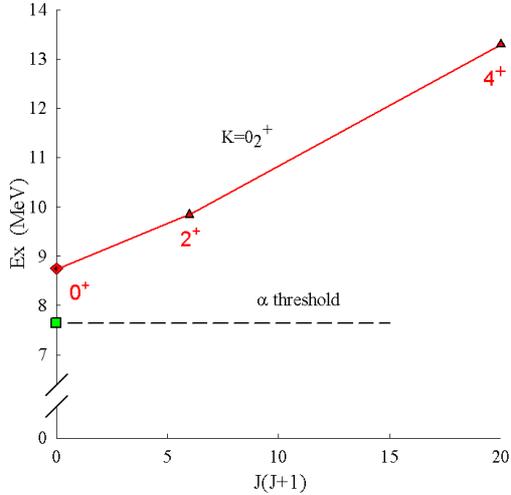}
 \protect\caption{\label{fig.3} {The experimental  states   with  the $\alpha$ 
 cluster  structure  in $^{12}$C:  the $0_3^+$ state  \cite{Itoh2012} (red diamond)
 of the Nambu-Goldstone boson,  the $2^+$ \cite{Itoh2011} and $4^+$  \cite{Freer2011}
 states (red triangle)  of the $K=0_2^+$   rotational band. The  $\alpha$
 condensate  Hoyle  state $0_2^+$ (green square) and the $0_4^+$ state (red circle)
 \cite{Itoh2012} are also  shown. The line is to guide the eye.
}
}
 \end{figure}

 In the literature no such a $0_3^+$ state with a zero-mode  nature
 has  been observed just above the Hoyle state. 
 The existence of  such a $0_3^+$ state will be important in the  nucleosynthesis of $^{12}$C 
from  triple $\alpha$ particle collisions. 
From this viewpoint, if we look into the excitation energy of the observed three states 
 carefully, the Hoyle state
 seems to be located slightly  below the energy expected by extrapolating  the 
rotational band of the $2_2^+$ (9.63 MeV) and $4_1^+$ (13.3 MeV)  states.
The estimated moment of inertia  connecting the Hoyle state and the $2_2^+$  (9.63 MeV) state
is about 80 \% of that estimated from the $2_2^+$ (9.63 MeV) and the $4_1^+$ (13.3 MeV)
  states.  This trend is  similar    to the
 four $\alpha$ case discussed before and  seems to suggest the  possible existence of a third
$0^+$ state from the  experimental side, because its existence slightly lowers  the position of the 
 Hoyle state due to quantum orthogonality. Therefore a third $0^+$ state is expected to 
exist slightly above the position extrapolated from the $J(J+1)$ line of the  rotational band for 
  the $2_2^+$ (9.63 MeV) and $4_2^+$ (13.3 MeV) states.  
If we look into the experimental data of the 
 isoscalar strength distribution of $^{12}$C($\alpha$,$\alpha$')
   in Fig.~8(a) of Ref.\cite{Itoh2011}, there is a peak in the excitation energy at
 around 8-9 MeV. This  seems to suggest the existence of a new 0$_3^+$ state at 
around 9 MeV 
just above the Hoyle state.  Very recently Itoh  et al.  \cite{Itoh2012} 
 observed  the  $0_3^+$ and $0_4^+$ states  with an $\alpha$
 cluster structure at 9.04 MeV (width 1.45 MeV) and 10.56 MeV (width 1.42 MeV),
 respectively. 
The existence of the $0_3^+$ state just above the Hoyle state on which a $K=0_2^+$  band is built
has logically the same  structure as  seen in the $^{16}$O case, which gives  strong support 
to  the $0_3^+$  state being  a NG boson collective mode state. The emergence of the NG boson
is   strong evidence  that  the Hoyle state is  a condensate of three $\alpha$ particles.
 The structure of Fig.~3  resembles  the $^{16}$O 
case in Fig.~1(b).
It is interesting to explore whether a  higher spin state built on the $0_4^+$, 
 which  could be a candidate for the band  corresponding to $\pi$(1300) in  Fig.~2,
 exists. Kurokawa and Kat\=o   \cite{Kurokawa2007} predict a $2^+$ state on the observed 
$0_4^+$ \cite{Itoh2012},  which  forms a $K=0_3^+$ band with almost the same moment of 
inertia as the $K=0_2^+$ band.

 We note  that  a dilute gaseous property  with a large radius is not 
  evidence for  condensation of $\alpha$ particles  without  firm evidence 
 such as superfluidity
 (superconductivity), a vortex or a  NG boson. Even  a non-condensate state
 has a gas-like dilute density distribution due to the threshold effect. 
For example, the 3$_1^-$ state in $^{12}$C, which is not a well-developed $\alpha$ cluster state,
 has a large radius comparable to the Hoyle state due to the threshold effect 
\cite{Danilov2009}. The $0_3^+$,  which is interpreted to have  
 a dominant [$^8$Be($2^+$)$\otimes\ell$=2]$_{J=0}$ structure in Ref.\cite{Uegaki1977}  and a linear
 chain-like structure  in Ref.\cite{Kanada2007}, also has  a much  larger radius 
than the Hoyle state. 
The dilute property  of a gaseous state of an $\alpha$ particle system arises for  the
 following four reasons: (1) the quantum $\alpha$ particle  condensation in momentum 
space, which brings about spatial  diluteness due to the uncertainty principle, (2) the NG boson
state, which causes diluteness because of its long wavelength nature, (3) the threshold 
effect by which the wave function extends near and above the Coulomb 
barrier and  (4) the Higgs boson state,
  which appears above the NG boson state, and  can be  dilute because of the collectivity
  of  the order parameter $\mid$$\Psi$$\mid$ and the threshold effect.
The present  mechanism is  general and 
the observation of a  NG boson collective 
mode with long wavelength (large moment of inertia) will be  useful  for the confirmation  
of an $\alpha$ particle condensate  near the  threshold energy.

To summarize,
we have investigated   $\alpha$ particle condensation in $^{16}$O and $^{12}$C 
from the viewpoint that  evidence of $\alpha$ particle condensation may be provided by 
the observation of the  NG boson caused by the spontaneous symmetry breaking
 of the global phase symmetry of a condensate. 
In $^{16}$O the $0^+$ state at 15.1 MeV is shown to be a condensate because 
the band head $0^+$ state at 16.6 MeV just above the 15.1 MeV $0^+$ state  is considered 
 to be a NG boson zero mode state.  The  $K=0_2^+$ band  built on top of it
 has a loosely coupled $\alpha$+$^{12}$C($0_2^+$) cluster structure corresponding
 well with  experimental observation. In $^{12}$C we showed  that the  newly observed
 $2_2^+$ (9.63 MeV) and $4_2^+$ (13.3 MeV) $\alpha$ cluster states above the Hoyle state are  
considered to be a rotational  band built on the NG boson $0^+$ state. This emerges
as a logical consequence of the Hoyle state being  a three $\alpha$ condensate state.
 It is interesting to investigate theoretically and experimentally  the emergence of the $\alpha$
 condensate  and the associated  Nambu-Goldstone  collective mode   in other nuclei. 

 The author thanks M.~Itoh for useful communications.

\end{document}